\documentclass[aps,prl,floatfix,showpacs,preprintnumbers,amsymb,a4paper,superscriptaddress,twocolumn]{revtex4-1}
\usepackage{amsmath}
\usepackage{graphicx}
\usepackage{amssymb}
\usepackage{amsmath}
\usepackage{epstopdf}
\usepackage{float}
\usepackage{color}

\newcommand{\be}{\begin{equation}}
\newcommand{\ee}{\end{equation}}

\begin{document}

\title{Multichannel long-range Rydberg molecules}
\author{Matthew T. Eiles}
\author{Chris H. Greene}
\date{\today }

\begin{abstract}
A generalized class of ultra-long-range Rydberg molecules is proposed which
consist of a multichannel Rydberg atom whose outermost electron creates a
chemical bond with a distant ground state atom. Such multichannel Rydberg
molecules exhibit favorable properties for laser excitation, because states
exist where the quantum defect varies strongly with the principal quantum
number. The resulting occurrence of near degeneracies with states of high
orbital angular momentum promotes the admixture of low $l$ into the high $l$
deeply bound `trilobite' molecule states, thereby circumventing the usual
difficulty posed by electric dipole selection rules. Such states also can
exhibit multi-scale binding possibilities that could present novel options
for quantum manipulation.
\end{abstract}

\maketitle





\affiliation{Department of Physics and Astronomy, Purdue University}





Ultra-long-range Rydberg molecules were predicted some time ago for the
simplest atoms in the periodic table, consisting of one ground state atom
and one Rydberg atom having just one valence electron outside of a
closed-shell core \cite{FermiOmont, Masnou-Seeuws, Lebedev, DuGreene87,
GreeneSadeghpourDickinson}. These unusual long-range Rydberg molecules have
since been observed experimentally, in a number of experiments that have
focused primarily on penetrating Rydberg states of low orbital angular
momentum. Very recently, experimental evidence \cite{Shaffer} of a true
long-range trilobite molecule has been found, consisting of an $ns$
Rydberg state of Cs with a large admixture of high angular momentum states
that hybridize to form an electronic wavefunction with a huge kilodebye
electric dipole moment.

The simplest type of such excited-ground state molecules have thus been
confirmed, and the theoretical description has been generalized to include
hyperfine interactions and the coupling of electronic singlet and triplet
states \cite{AndersonRaithel}. Moreover, other interesting generalizations
have been developed, such as the binding of multiple ground state atoms\cite%
{Rost2006, Pfau}, the behavior of long-range Rydberg molecules subjected to
an external electric field\cite{dePrunele, HamiltonThesis, KurzSchmelcher},
and even the behavior of a single Rydberg alkali atom excited in a
Bose-Einstein condensate \cite{PfauBEC}.

While an alkali atom Rydberg gas provides a simple and easily-controlled
system for the excitation of novel molecular states, in fact Rydberg atoms
throughout the rest of the periodic table are far richer and more complex,
as they consist primarily of perturbed, multichannel Rydberg spectra \cite%
{FanoJOSA,AymarReview1984,OrangeRMP,GallagherBook}. The present Letter
demonstrates some of the opportunities provided by this richer class of
multichannel Rydberg atoms that can bind one or more ground state atoms. One
finding in particular is that the existence of doubly-excited perturbers can
be used in the alkaline earth atoms to directly excite long-range trilobite
molecules with their exceptionally large electric dipole moments. Another
possibility raised by the existence of doubly-excited perturbers is that
energy eigenfunctions exist whose multichannel character exhibits two very
different length scales. Such stationary states can in principle be used to
trap ground state atoms at very different internuclear distances, providing
opportunities for unusual types of quantum control and manipulation.

\begin{figure}[h]
\includegraphics[scale = 0.3]{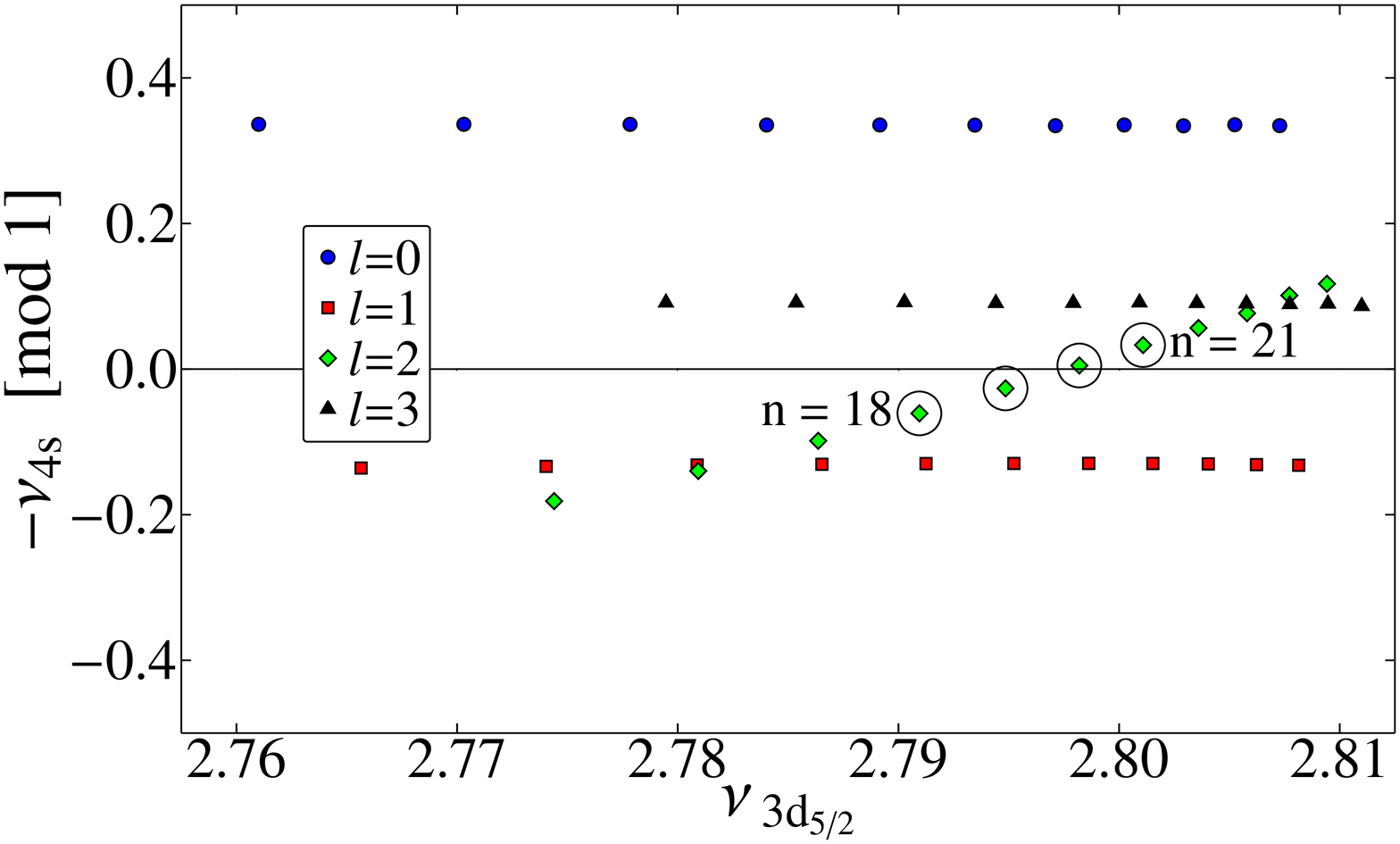}
\caption{{A Lu-Fano plot \protect\cite{LuFano} highlighting the effects of
level perturbers on the quantum defects of the $nd$ states of Ca 
\protect\cite{Nist}, in contrast to the energy-independent behavior of the
quantum defects for $s$, $p$, and $f$ states. The quantum number $n$ rises
from 15 to 24 from left to right. This letter investigates the circled
states. }}
\label{energydiagram}
\end{figure}
This letter focuses on the 4snd $^1D_2$ levels in calcium, where
perturbations from the doubly excited states 3d4d and 3d5s $^1D_2$ cause the
4snd singlet quantum defect to vary rapidly in the vicinity of these
perturbing levels \cite{AymarTelmini1991}. Fig. \ref{energydiagram}
depicts the strong $n$-dependence of the $l = 2$ quantum defects. Calcium is
chosen because the d-wave quantum defects pass through unity as the
principal quantum number $n$ increases between $n \sim 20\pm 2$; the $nd$
state is therefore highly degenerate with high angular momentum states in
the $n-1$ manifold. This accidental degeneracy strongly couples these states
together, allowing for direct two-photon excitation of trilobite molecules,
circumventing the usual challenges of exciting high angular momentum states
which typically require electric fields to break dipole selection rules or
else an additional microwave photon after a two-photon process \cite%
{GreeneSadeghpourDickinson}. This removes the experimental obstacles in
creating and studying these exotic long-range molecules.
The interaction between the nearly-free Rydberg electron and the ground
state perturber can be simply and accurately described by the Fermi
pseudo-potential \cite{FermiOmont} 
\begin{equation}
\begin{split}
{\small V_{pp}(\vec{r},\vec{R})}& {\small =2\pi a_{s}[k(R)]\delta ^{3}(\vec{r%
}-R\hat{z})} \\
& \,\,\,\,\,\,\,+6\pi a_{p}[k(R)]\delta ^{3}(\vec{r}-R\hat{z})%
\reflectbox{\ensuremath{\vec{\reflectbox{\ensuremath{\nabla}}}}}\cdot \vec{%
\nabla},
\end{split}%
\end{equation}%
where the energy-dependent s-wave scattering length and p-wave scattering
volume are respectively, {\small $a_{s}[k(R)]=-\tan \delta _{0}/k(R)$ }and%
{\normalsize \ }{\small $a_{p}[k(R)]=-\tan \delta _{1}/[k(R)]^{3}$ },
implicitly depend on the internuclear separation $R$ through the
semiclassical kinetic energy $\frac{1}{2}k(R)^{2}=E_{nl}+\frac{1}{R}$. The
internuclear axis points in the $\hat{z}$ direction. The energy-dependent
phase shift data used to calculate these scattering properties were
calculated by \cite{BartschatSadeghpour}. Both earlier \cite{DuGreene87} and
recent \cite{Fey} work has emphasized that diagonalization using this
delta-function pseudopotential is formally non-convergent, but comparison
with more complicated Green function or Kirchhoff integral methods \cite%
{KhuskivadzeChibisovFabrikant,HamiltonThesis} shows that nevertheless the
models agree to first order. Quantitative accuracy with experimental data
within the pseudopotential model can be achieved by fitting the scattering
lengths to observed spectra \cite{Bendkowsky}.

Two characteristics of the phase shifts differentiate this interaction in
calcium from those previously studied in the alkalis. Since calcium does not
possess the p-wave shape resonance that significantly modifies the potential
curves of the alkali dimers \cite{BahrimThumm,BartschatSadeghpour}, the
potential wells at all but very small internuclear distances are determined
almost entirely by s-wave electron-atom scattering properties. The second
contrast is that the p-wave phase shift is negative except at very low
energies, so the scattering volume is positive over nearly the entire range
of internuclear distances. These two differences eliminate the strongly
attractive p-wave interactions present in the alkalis that support
\textquotedblleft butterfly\textquotedblright\ type bound states \cite%
{HamiltonGreeneSadeghpour}, and instead the p-wave interaction produces a
weakly repulsive potential. Bound states could possibly form in this
repulsive potential in the presence of a second perturber \cite{Rost2009}. Ca$^-$ also exists as a bound negative ion with binding of the doublet $P$ electron by $24.55\pm10$ meV for $J=1/2$ and by $19.73 \pm 10$ meV for $J=3/2$, whereas in triplet e-Rb and e-Cs there is no stable negative ion. \cite{TAnderson}

\begin{figure*}[tbp]
{\normalsize 
\includegraphics[scale =
0.2]{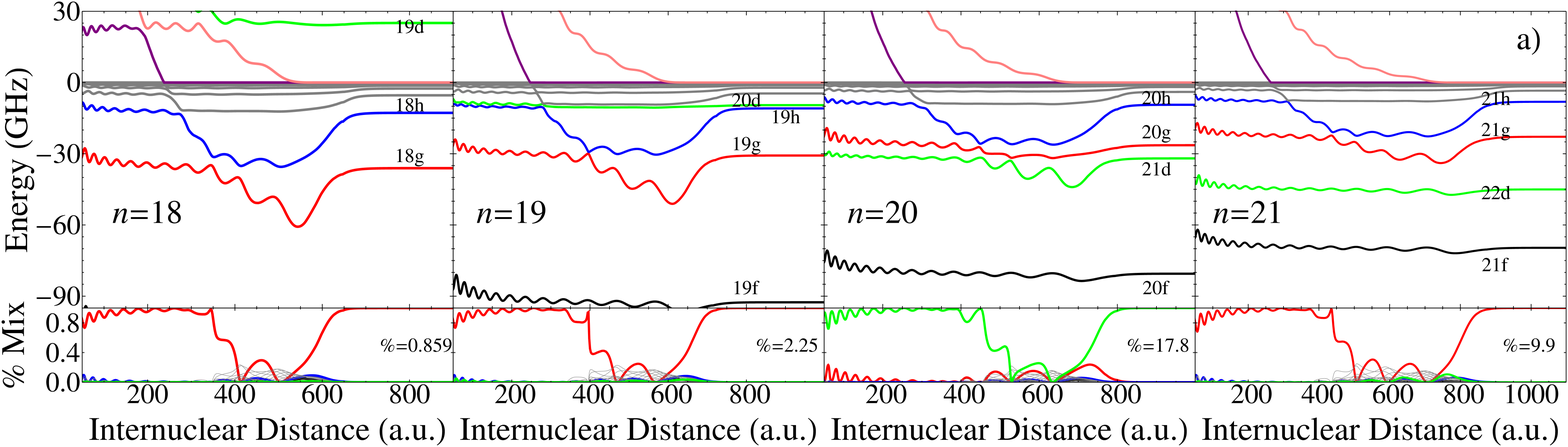}\newline
\vspace{-10pt} 
\includegraphics[scale =
0.2]{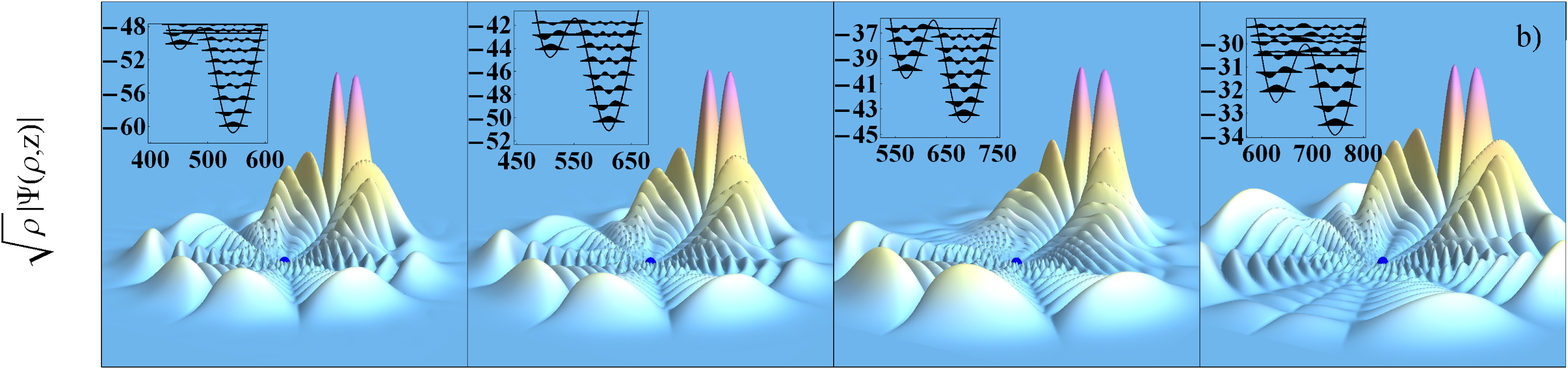}\newline
\vspace{-20pt} 
\includegraphics[scale =
0.2]{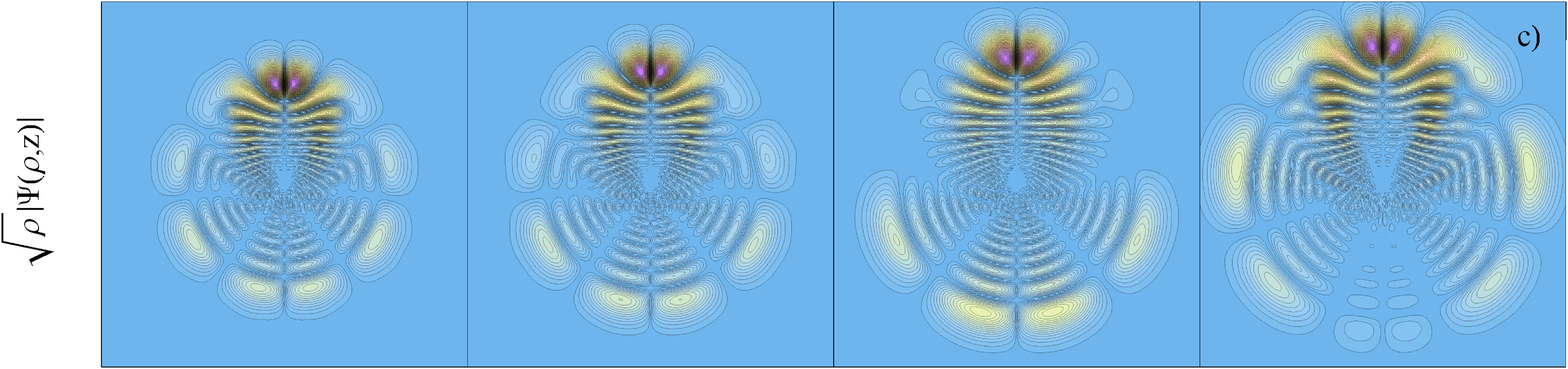}\newline
}
\caption{a) Potential curves for $\Omega =0$ states, with $n$ increasing
from 18-21 from left to right. The energy scale is relative to $-\frac{1}{%
2n^{2}}$. The $(n+1)d$ state descends through the degenerate manifold with
increasing $n$. The expansion coefficients are plotted with the same color
scheme, showing the locations of the trilobite states (large mixing of high
angular momentum states) and importantly the amount of $(n+1)d$ mixing in
the trilobite state. The percentage labeled
gives the percent contribution of the d state in the trilobites plotted in panels b) and c).\newline
b) Electron probability amplitudes $\protect\sqrt{\protect\rho }|\Psi (%
\protect\rho ,z)|$ in cylindrical coordinates over a region in the $(\protect%
\rho ,z,0)$ plane spanning $(-1000,1000)$. This wavefunction is in the
deepest minimum of the potential wells, shown in the inset along with the
lowest vibrational states (in GHz, as a function of internuclear distance) of the Rydberg molecule. The Rydberg nucleus is
marked by the blue sphere at the origin; the ground state atom is located in
between the tallest peaks of the wavefunction at $z\sim 600\rightarrow
800a_{0}$.\newline
c) Contour plots of the probability amplitudes, showing the transition from
primarily f-wave character outside of the trilobite wavefunction in $n=18$, $%
19$ to d-wave character in $20$. The scale is the same in all four graphs.}
\label{bigplot}
\end{figure*}
Adiabatic potential energy curves for $18\leq n\leq 21$ calculated by basis
set diagonalization are presented in Fig. (\ref{bigplot}a). As $n$
increases the $(n+1)d$ state sweeps downwards through the $n$ manifold, and
for $n=19,20$ in particular it is strongly mixed into the trilobite state. \
This systematic change of quantum defects with energy, caused by familiar
level perturbations in multichannel spectroscopy\cite{OrangeRMP}, is
ubiquitous in the heavier alkaline earth metal atoms and in most other atoms
in the periodic table. \cite{GreeneSadeghpourDickinson} predicts two classes
of long-range Rb molecules: low-angular momentum states (a) that have been
observed experimentally to admix a fractional amount of trilobite character
giving them non-negligible permanent electric dipole moments \cite{Li}, and
(b) the trilobite molecules themselves, consisting of an admixture of high
angular momentum states. The states shown here in Ca have a level of mixing
intermediate between these two regimes, akin to the recent results in Cs 
\cite{Shaffer}. Typical trilobite electronic probability amplitudes in
cylindrical coordinates displayed in Fig. (\ref{bigplot}b,c) demonstrate
this mixing as they show a \textquotedblleft trilobite\textquotedblright\
wavefunction superimposed over an $f$ or $d$-state wavefunction.

The molecular vibrational levels seen in the potential well inset are
approximately spaced at 1 GHz. Many vibrational bound states are supported
in these wells, which range from 30 to 60 GHz deep. These molecules should
therefore be somewhat more tightly bound than those predicted for rubidium 
\cite{GreeneSadeghpourDickinson,KhuskivadzeChibisovFabrikant}. The permanent
electron dipole moments associated with these molecules are in the kdebye
range, allowing for the possibility of delicate manipulation of these
molecules with electric and magnetic fields \cite{KurzMayle,Lesanovsky}.

\begin{figure}[tbp]
{\normalsize \includegraphics[scale = 0.45]{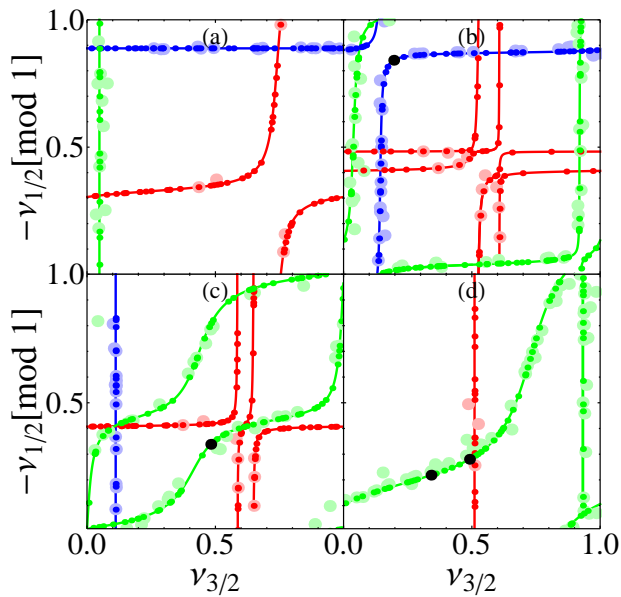} 
}
\caption{Lu-Fano plots for a) $J=0$, b) $J=1$, c) $J=2$, and d) $J=3$
symmetries. Blue points are $l_e\approx 0$ odd parity; red are $l_e \approx 1$ even parity, and green are $l_e \approx 2$ odd parity.
Larger points represent experimental levels; smaller points theoretically
predicted. Black points mark the states in Fig (\ref{MQDTBOPECS})}
\label{lufanoJs}
\end{figure}

A second example of the rich physics provided by two-electron systems is
predicted by exploiting the complex channel interactions due to the
fine-structure splitting to create long-range molecules that display two
distinct length scales. In silicon the $3p$ $^{2}P_{1/2}^{o}$ and $3p$ $%
^{2}P_{3/2}^{o}$ ionization thresholds are separated by $\Delta E=287.84\,%
\text{cm}^{-1}$. Nonperturbative interactions between Rydberg series in
different channels converging to these two thresholds require the framework
of multichannel quantum defect theory (MQDT) to describe the bound states of
the atom. Silicon has been studied in this framework both from a
semi-empirical standpoint \cite{BrownGinterGinter} and through nearly ab
initio R-matrix calculations \cite{GreeneKim,FrancisGreene,OrangeRMP}. Only
the salient details of the semi-empirical approach will be described here,
since the references contain thorough descriptions \cite%
{Seaton,OrangeRMP,CookeCromer,FanoRau}. The elements of a diagonal
short-range reaction matrix in $LS$ coupling, $K_{ii^{\prime
}}^{(LS)}=\delta _{ii^{\prime }}\tan \pi \mu _{i}$ are treated as fit
parameters by matching experimental bound states with MQDT predictions. The
advantage of the $K$-matrix formulation is that these elements have slow
energy dependence in the short-range region near the core since asymptotic
boundary conditions have not yet been enforced. The $i^{\prime }-$th
linearly independent wavefunction outside of this region is 
\begin{equation}
\Psi _{i^{\prime }}=\mathcal{A}\sum_{i}\Phi _{i}(\Omega )\left[
f_{i}(r)\delta _{ii^{\prime }}-g_{i}(r)K_{ii^{\prime }}^{(LS)}\right] ,
\end{equation}%
where $i$ labels different channels, $\Phi _{i}(\Omega )$ contains the
wavefunction of the atomic core and all spin and orbital angular momentum
degrees of freedom of the Rydberg electron, and $(f,g)$ are the
regular/irregular solutions to the Schr\"{o}dinger equation with a Coulomb
potential outside of the atomic core \cite{FanoRau}. $\ $An
antisymmetrization operator is denoted here as $\mathcal{A}.$ \ LS coupling,
which couples the orbital and spin angular momenta separately and is
described by the ket $|[(l_{c}l_{e})L(\frac{1}{2}\frac{1}{2})S]JM_{J}\rangle 
$ is accurate for low-lying states where the electron is near the core and
exchange effects dominate \cite{FanoRau}. This coupling scheme breaks down
at long-range where the electron accumulates radial phase at rates strongly
dependent on the atomic core's total angular momentum \cite{OrangeRMP}. Here
a \textquotedblleft geometric\textquotedblright\ orthogonal frame
transformation matrix $U_{ij}$, given by standard angular momentum algebra,
is used to transform into the more appropriate $jj$-coupling scheme
represented by the ket $|[(l_{c}\frac{1}{2})J_{c}(l_{e}\frac{1}{2}%
)J_{e}]JM_{J}\rangle $ \cite{LeeLu}. The $jj$-coupled $K$ matrix is $%
K_{ii^{\prime }}^{(jj)}=\sum_{jj'}U_{ij}K_{jj^{\prime }}^{(LS)}U_{j^{\prime }i^{\prime
}}^{T }$. Imposition of boundary conditions at long-range requires
that 
\begin{equation}
\left( \delta _{ii^{\prime }}\sin \beta _{i}+\cos \beta _{i}K_{ii^{\prime
}}^{(jj)}\right) B_{i^{\prime }}=0;\,\,\beta _{i}=\pi (\nu _{i}-l_{i}).
\end{equation}%
Ensuring a vanishing determinant constrains the allowed values of $\nu _{i}$
to take on discrete values, and the eigenvector $B_{i}$ can be matched at
long-range to an expansion in terms of re-normalized Whittaker functions $%
W(r,\nu _{i},l_{i})$ that exponentially decay at long-range (\cite%
{OrangeRMP} eq. 2.53): 
\begin{equation}
\Psi _{i^{\prime }}={\cal A}\sum_{i}\left[ -\frac{B_{i}}{\cos {\beta _{i}}}\right]
\Phi _{i}(\Omega )W_{i}(r,\nu _{i},l_{i}).
\end{equation}%
The critical parameters in this expression are the mixing coefficients $%
B_{i}/\cos (\pi \beta _{i})$, which determine the weighting of each channel
eigenfunction in the energy eigenstate. These coefficients, as well as the
bound state energies, were determined here by fitting the quantum defects $%
\mu _{i}$ and a set of orthogonal rotation matrices to high-lying
experimental levels. Spectroscopic data for $l_{e}=0,2$ odd parity and $l_{e}=1$ even parity states with $J=0-3$ were fitted this way, with results (Fig. \ref{lufanoJs}) that compare favorably
with R-matrix methods \cite{GreeneKim,FrancisGreene}. Most high-lying
experimental energies were fitted to within $0.5\text{cm}^{-1}$, but the
results for $l_{e}=1$ are very uncertain due to a dearth of experimental
values available for fitting. The model's accuracy can be seen by comparing
the larger points on Fig. \ref{lufanoJs} with the theoretical points. 

\begin{figure}[tbp]
{\normalsize \includegraphics[scale = 0.35]{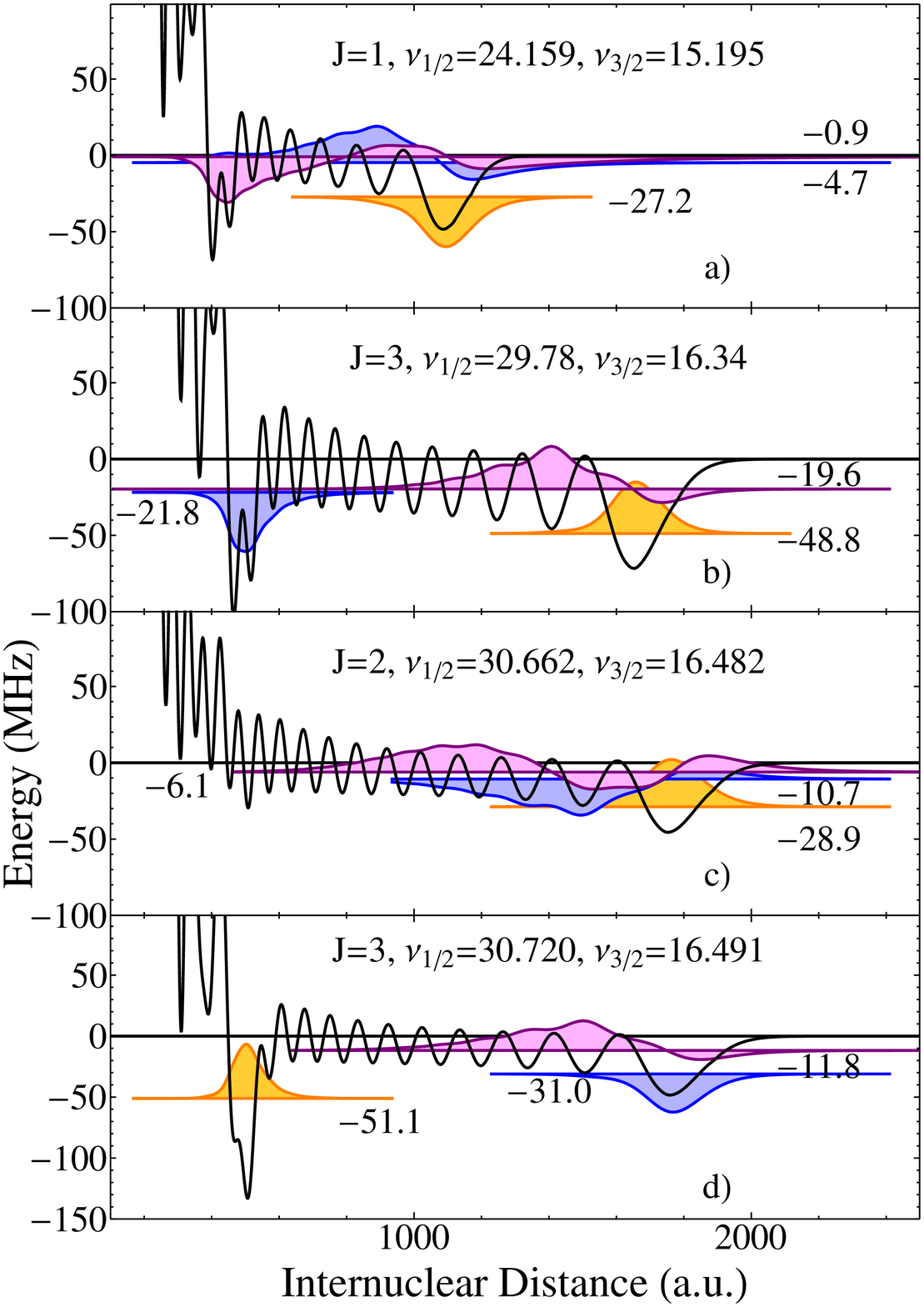} }
\caption{Adiabatic potential curves, bound state energies, and molecular
bound states for various low-$J$, levels of the $M_{J}=0$ Si$^{\ast }$-Ca
Rydberg molecule. The binding energies (in MHz) labeling each bound state are listed relative to the asymptotic atomic energy, which
can be obtained from $E = E_1-1/2\nu_{1/2}^2=E_2-1/2\nu_{3/2}^2$. We estimate an uncertainty of $0.5$ MHz in these values. } 
\label{MQDTBOPECS}
\end{figure}

The system studied here consists of a Rydberg Si atom interacting via the
Fermi pseudopotential with a dilute gas of Ca ground state atoms. We focus
on the low angular momentum potential curves, similar to class (a) from \cite%
{GreeneSadeghpourDickinson}. The p-wave contribution is non-negligible at small internuclear distances, where the potential as a result becomes repulsive. Fig. \ref{MQDTBOPECS} displays several example potential curves. Those states exhibiting the
multi-scale binding behavior predicted here lie on
rapidly varying portions of the Lu-Fano plot, while primarily single-channel
states fall on flat portions of these curves; hence, the $J = 3$, $l_{e}\approx
2$ states most consistently display deep separated wells. Panels b) and c) of Fig. (\ref{MQDTBOPECS}) show two typical potential curves for this symmetry, while a) shows one of the few $J = 1$, $l_e \approx 0$ states to display two well separated deep wells, and c) shows a primarily single channel case of the $J = 2$, $l_e \approx 2$ symmetry. The increasing well separation
is due to the spatial scaling of each single-channel wavefunction with $\nu
_{i}^{2}$. The potential curves are typically separated from
neighboring curves by several hundred MHz or more and each support a few bound states with binding energies typically smaller than those investigated in Rb experiments 
\cite{Bendkowsky,Pfau}. The ability of these
potentials to support bound states is enhanced by the relatively large
spin-orbit splitting and high molecular reduced mass of the Si-Ca system, but depends delicately on the energy and symmetry of the component states. 

These two scenarios illustrate the rich effects of the physics of long-range
multichannel molecules formed by divalent atoms. Interactions with doubly
excited states in Ca allow for direct excitation of trilobite molecules
through low-$l$ channels, providing a new method for experimental
realization of these molecules through direct excitation. This represents a
major reduction in the experimental challenges in forming these molecules,
and together with recent experimental and theoretical efforts with Sr \cite%
{Burgdorfer,Vaillant} suggest that the alkaline atoms will be fertile fields
of progress in ultralong-range molecular studies. In the Si$^{\ast }$-Ca
molecules predicted here, the multichannel description of the Rydberg
structure predicts far richer low angular momentum potential energy curves
than those studied in the alkalis, which formed bound states essentially
only in the outermost well. The highly localized states in these wells
should be spectroscopically accessible leading to possible applications in
manipulation of the ground state atom's position across hundreds of atomic
units.

\begin{acknowledgments}
We thank J. Perez-Rios and P. Giannakeas for many helpful discussions. This work is supported in part by the National Science Foundation under Grant No. PHY-1306905.
\end{acknowledgments}



\begin{thebibliography}{99}
\bibitem{GreeneSadeghpourDickinson} C. H. Greene, A. S. Dickinson and H. R.
Sadeghpour, Phys. Rev. Lett. {\bf 85}, 2458 (2000).

\bibitem{FermiOmont} E. Fermi, Nuovo Cimento {\bf 11}, 157 (1934); A. Omont, J.
Phys. (Paris) {\bf 38}, 1343 (1977).

\bibitem{Masnou-Seeuws} P. Valiron, A. L. Roche, F.
Masnou-Seeuws, and M. E. Dolan, J. Phys. B {\bf 17}, 2803 (1984).

\bibitem{Lebedev} I. L. Beigman and V. S. Lebedev, Phys. Rep. {\bf 250}, 95 (1995).

\bibitem{DuGreene87} N. Y. Du and C. H. Greene, Phys. Rev. A {\bf 36}, 971 (1987). See also the erratum, {\it ibid.} {\bf 36}, 5467 (1987).

\bibitem{Shaffer} D. Booth, S. T. Rittenhouse, J. Yang, H. R. Sadeghpour, \&
J. P. Shaffer, Science, {\bf 348}, 99 (2015).

\bibitem{AndersonRaithel} D. A. Anderson, S. A. Miller, and G. Raithel, Phys.
Rev. A {\bf 90}, 062518 (2014).

\bibitem{Rost2006} I. C. H. Liu and J. M. Rost, Eur. Phys. J. D {\bf 40}, 65
(2006).

\bibitem{Pfau} V. Bendkowsky, B. Butscher, J. Nipper, J.~B. Balewski, J. P.
Shaffer, R. L\"{o}w, T. Pfau, W. Li, J. Stanojevic, T. Pohl, and J. M. Rost,
Phys. Rev. Lett. {\bf 105}, 163201 (2010).

\bibitem{KurzSchmelcher} M. Kurz, and P. Schmelcher, Phys. Rev. A {\bf 88}, 022501
(2013).

\bibitem{HamiltonThesis} E. L. Hamilton, Ph.D. thesis, University of
Colorado, (2002).

\bibitem{dePrunele} E. de Prunel\'{e}, Phys. Rev. A {\bf 35}, 496 (1987).

\bibitem{PfauBEC} J. B. Balewski, A. T. Krupp, A. Gaj, D. Peter, H. P. B\"{u}%
chler, R. L\"{o}w, S. Hofferberth, and T. Pfau, Nature (London) {\bf 502}, 664
(2013).

\bibitem{OrangeRMP} M. Aymar, C. H. Greene, and E. Luc-Koenig, Rev. Mod.
Phys. {\bf 68}, 1015 (1996).

\bibitem{FanoJOSA} U. Fano, J. Opt. Soc. Am. {\bf 65}, 979 (1975).

\bibitem{AymarReview1984} M. Aymar, Phys. Rep. {\bf 110}, 163 (1984).

\bibitem{GallagherBook} T. F. Gallagher, {\it Rydberg Atoms} (Cambridge University Press, Cambridge, England, 2005)

\bibitem{AymarTelmini1991} M. Aymar and M. Telmini, J. Phys. B: At. Mol.
Opt. Phys. {\bf 24}, 4935 (1991).

\bibitem{LuFano} K. T. Lu and U. Fano, Phys. Rev. A {\bf 2}, 81 (1970).

\bibitem{Nist} J. Sugar and C. Corliss, J. Phys. Chem. Ref. Data {\bf 14}, Suppl. No. 2, 1 (1985).

\bibitem{Rost2009} I. C. H. Liu, J. Stanojevic, and J. M. Rost, Phys.
Rev. Lett. {\bf 102}, 173001 (2009).

\bibitem{BartschatSadeghpour} K. Bartschat and H. R. Sadeghpour, J. Phys. B {\bf 36}, L9 (2003); J. Yuan and Z. Zhang, Phys. Rev. A {\bf 42},
5363 (1990).

\bibitem{KhuskivadzeChibisovFabrikant} A. A. Khuskivadze, M. I. Chibisov,
and I. I. Fabrikant, Phys. Rev. A {\bf 66}, 042709 (2002); M. I. Chibisov, A. A.
Khuskivadze, and I. I. Fabrikant, J. Phys. B  {\bf 35}, L193 (2002).

\bibitem{Fey} C. Fey, M. Kurz, P. Schmelcher, S. T. Rittenhouse, and H. R.
Sadeghpour, New J. Phys. {\bf 17} 055010 (2015).

\bibitem{Bendkowsky} V. Bendkowsky, B. Butscher, J. Nipper, J. B. Balewski,
J. P. Shaffer, R. L\"{o}w, T. Pfau, Nature (Londong) {\bf 458} 1005 (2009).

\bibitem{BahrimThumm} C. Bahrim, U. Thumm and I.I. Fabrikant, Phys. Rev. A
{\bf 63} 042710 (2001).

\bibitem{HamiltonGreeneSadeghpour} E. L. Hamilton, C. H. Greene, and H. R.
Sadeghpour, J. Phys. B {\bf 35}, L199 (2002).

\bibitem{DuGreene89} N. Y. Du and C. H. Greene, J. Chem. Phys. {\bf 90}, 6347
(1989). 

\bibitem{TAnderson} T. Andersen, Phys. Rep. {\bf 394}, 157 (2004).

\bibitem{Li} W. Li, T. Pohl, J. M. Rost, S.~T. Rittenhouse, H.~R.
Sadeghpour, J. Nipper, B. Butscher, J.~B. Balewski, V. Bendkowsky, R. L\"{o}%
w, and T. Pfau, Science {\bf 334}, 1110 (2011).

\bibitem{KurzMayle} M. Kurz, M. Mayle, and P. Schmelcher, Europhys. Lett. {\bf 97}, 43001
(2012).

\bibitem{Lesanovsky} I. Lesanovsky, P. Schmelcher, and H. R. Sadeghpour, J.
Phys. B {\bf 39}, L69 (2006).

\bibitem{GonzalezFerez} R. Gonz\'{a}lez-F\'{e}rez, H. R. Sadeghpour, and P.
Schmelcher, New J. Phys. {\bf 17} 013021 (2015).

\bibitem{BrownGinterGinter} C. M. Brown, S. G. Tilford, and M. L. Ginter, J.
Opt. Soc. Am. {\bf 65}, 385 (1975); D. S. Ginter, M. L. Ginter, and C. M. Brown,
J. Chem. Phys. {\bf 85}, 6530 (1986); D. S. Ginter and M. L. Ginter, J. Chem. Phys
{\bf 85}, 6536 (1986).

\bibitem{FrancisGreene} F. Robicheaux and C. H. Greene, Phys. Rev. A {\bf 47},
4908 (1993).

\bibitem{GreeneKim} L. Kim and C. H. Greene, Phys. Rev. A {\bf 36}, 4272 (1987);
{\bf 38}, 5953 (1988).

\bibitem{LeeLu} C. M. Lee and K. T. Lu, Phys. Rev. A {\bf 8}, 1241 (1973); C. H.
Greene, J. Opt. Soc. Am. B {\bf 4}, 775 (1987).

\bibitem{Seaton} M. J. Seaton, Rep. Prog. Phys. {\bf 46}, 167 (1983), and
references therein.

\bibitem{CookeCromer} W. E. Cooke and C. L. Cromer, Phys. Rev. A {\bf 32}, 2725
(1985).

\bibitem{FanoRau} U. Fano and A. R. P. Rau, {\it Atomic Collisions and Spectra}
(Academic, Orlando, 1986).

\bibitem{Burgdorfer} B. J. DeSalvo, J. A. Aman, F. B. Dunning, T. C. Killian,
H. R. Sadeghpour, S. Yoshida, J. Burgd\"{o}rfer, arXiv:1503.07929 (2015).

\bibitem{Vaillant} C. L. Vaillant, M. P. A. Jones, and R. M. Potvliege, J.
Phys. B  {\bf 47}, 155001 (2014).
\end{thebibliography}
\end{document}